\documentclass[10pt]{article}

\usepackage{amsmath}
\usepackage{amssymb}

\usepackage{graphicx}

\usepackage{cite}

\usepackage{lineno}

\usepackage[labelfont=bf,labelsep=period,justification=raggedright]{caption}

\makeatletter
\renewcommand{\@biblabel}[1]{\quad#1.}
\makeatother

\date{}

\begin{document}

% Title must be 150 characters or less
\begin{flushleft}
{\Large
\textbf{Measuring integrated information from the decoding perspective}
}
% Insert Author names, affiliations and corresponding author email.
\\
Masafumi Oizumi$^{1,2,\ast}$, 
Shun-ichi Amari$^{1}$,
Toru Yanagawa$^{1}$,
Naotaka Fujii$^{1}$,
Naotsugu Tsuchiya$^{2,3,\ast}$
\\
\bf{1} RIKEN Brain Science Institute, 2-1 Hirosawa, Wako, Saitama 351-0198, Japan
\\
\bf{2} Monash University, Clayton Campus, Victoria 3800, Australia
\\
\bf{3} Japan Science and Technology Agency, Japan
\\
$\ast$ E-mail: oizumi@brain.riken.jp, naotsugu.tsuchiya@monash.edu
\end{flushleft}

% Please keep the abstract between 250 and 300 words
\section*{Abstract}
Accumulating evidence indicates that the capacity to integrate information in the brain is a prerequisite for consciousness. Integrated Information Theory (IIT) of consciousness provides a mathematical approach to quantifying the information integrated in a system, called integrated information, $\Phi$. Integrated information is defined theoretically as the amount of information a system generates as a whole, above and beyond the sum of the amount of information its parts independently generate. IIT predicts that the amount of integrated information in the brain should reflect levels of consciousness. Empirical evaluation of this theory requires computing integrated information from neural data acquired from experiments, although difficulties with using the original measure $\Phi$ precludes such computations. Although some practical measures have been previously proposed, we found that these measures fail to satisfy the theoretical requirements as a measure of integrated information. Measures of integrated information should satisfy the lower and upper bounds as follows: The lower bound of integrated information should be 0 when the system does not generate information (no information) or when the system comprises independent parts (no integration). The upper bound of integrated information is the amount of information generated by the whole system and is realized when the amount of information generated independently by its parts equals to 0. Here we derive the novel practical measure $\Phi^*$ by introducing a concept of mismatched decoding developed from information theory. We show that $\Phi^*$ is properly bounded from below and above, as required, as a measure of integrated information. We derive the analytical expression $\Phi^*$ under the Gaussian assumption, which makes it readily applicable to experimental data. Our novel measure $\Phi^*$ can be generally used as a measure of integrated information in research on consciousness, and also as a tool for network analysis in research on diverse areas of biology.

% Please keep the Author Summary between 150 and 200 words
% Use first person. PLOS ONE authors please skip this step.
% Author Summary not valid for PLOS ONE submissions.
\section*{Author Summary}
Integrated Information Theory (IIT) of consciousness attracts scientists who investigate consciousness owing to its explanatory and predictive powers for understanding the neural properties of consciousness. IIT predicts that the levels of consciousness are related to the quantity of information integrated in the brain, which is called integrated information $\Phi$. Integrated information measures excess information generated by a system as a whole above and beyond the amount of information independently generated by its parts. Although IIT predictions are indirectly supported by numerous experiments, validation is required through quantifying integrated information directly from experimental neural data. Practical difficulties account for the absence of direct, quantitative support. To resolve these difficulties, several practical measures of integrated information have been proposed. However, we found that these measures do not satisfy the theoretical requirements of integrated information: first, integrated information should not be below 0; and second, integrated information should not exceed the quantity of information generated by the whole system. 

Here, we propose a novel practical measure of integrated information, designated as $\Phi^*$ that satisfies these theoretical requirements by introducing the concept of mismatched decoding developed from information theory. $\Phi^*$ creates the possibility of empirical and quantitative validations of IIT to gain novel insights into the neural basis of consciousness.

\section*{Introduction}
Although its neurobiological basis remains unclear, consciousness may be related to certain aspects of information processing \cite{Chalmers1995, Tononi2004}. In particular, Integrated Information Theory of consciousness (IIT) developed by Tononi and colleagues \cite{Tononi2004, Tononi2008, Tononi2010, Tononi2012, Balduzzi2008, Balduzzi2009, Oizumi2014, Tononi2015} predicts that the amount of information integrated among the components of a system, called integrated information $\Phi$, is related to the level of consciousness of the system. The level of consciousness in the brain varies from a very high level, as in full wakefulness, to a very low level, as in deeply anesthetized states or dreamless sleep. When consciousness changes from high to low, IIT predicts that the amount of integrated information changes from high to low, accordingly. This prediction is indirectly supported by recent neuroimaging experiments that combine noninvasive magnetic stimulation of the brain (transcranial magnetic stimulation, TMS) with electrophysiological recordings of stimulation-evoked activity (electroencephalography) \cite{Massimini2005, Massimini2007, Ferrarelli2010, Rosanova2012, Casali2013}. Such evidence implies that if there is a practical method to estimate the amount of integrated information from neural activities, we may be able to measure levels of consciousness using integrated information.

IIT provides several versions of mathematical formulations to calculate integrated information \cite{Tononi2004, Tononi2008, Tononi2010, Tononi2012, Balduzzi2008, Balduzzi2009, Oizumi2014}. Although the detailed mathematical formulations are different, the central philosophy of integrated information does not vary among different versions of IIT. Integrated information is mathematically defined as the amount of information generated by a system as a whole above and beyond the amount of information generated independently by its parts. If the parts are independent, integrated information will not exist.

Despite its potential importance, the empirical calculation of integrated information is difficult. For example, one difficulty involves making an assumption when integrated information is calculated according to the informational relationship between past and present states of a system. The distribution of past states is assumed to maximize entropy, which is called the maximum entropy distribution. The assumption of maximum entropy distribution severely limits the applicability of the original integrated information measure $\Phi$ indicated by \cite{Barrett2011}. First, the concept of maximum entropy distribution cannot be applied to a system that comprises elements whose states are continuous, because there is no unique maximum entropy distribution for continuous variables \cite{Cover1991, Barrett2011}. Second, information under the assumption of the maximum entropy distribution can be computed only when there is complete knowledge about the transition probability matrix that describes how the system transits between states. However, the transition probability matrix for actual neuronal systems is practically impossible to estimate for all possible states. 

To overcome these problems, Barrett and Seth \cite{Barrett2011} proposed using the empirical distribution estimated from experimental data, thereby removing the requirement to rely on the assumption of the maximum entropy distribution. Although we believe that their approach does lead to practical computation of integrated information, we found that their proposed measures based on empirical distribution \cite{Barrett2011} do not satisfy key theoretical requirements as a measure of integrated information. Two theoretical requirements should be satisfied as a measure of integrated information. First, the amount of integrated information should not be negative. Second, the amount of integrated information should never exceed information generated by the whole system. These theoretical requirements, which are satisfied by the original measure $\Phi$, are required so that a measure of integrated information is interpretable in accordance with the original philosophy of integrated information, i.e., integrated information measures the extra information generated by a system as a whole above and beyond the amount of information independently generated by its parts. 
 
Here, we propose a novel practical measure of integrated information, $\Phi^*$, by introducing the concept of mismatched decoding developed from information theory \cite{Merhav1994, Latham2005, Oizumi2010}. $\Phi^*$ represents the difference between ``actual" and ``hypothetical'' mutual information between past and present states of the system. The actual mutual information corresponds to the amount of information that can be extracted about past states by knowing present states (or vice versa) when the actual probability distribution of a system is used for decoding information for past and present states. In contrast, hypothetical mutual information corresponds to the amount of information that can be extracted about past states by knowing present states when a ``mismatched'' probability distribution is used for decoding where a system is partitioned into hypothetical independent parts. Decoding with a mismatched probability distribution is called mismatched decoding. $\Phi^*$ quantifies the amount of loss of information caused by the mismatched decoding where interactions between the parts are ignored. We show here that $\Phi^*$ satisfies the theoretical requirements as a measure of integrated information, unlike the previously proposed measures. Further, we derive the analytical expression of $\Phi^*$ under the Gaussian assumption and make this measure feasible for practical computation.

% Results and Discussion can be combined.
% We only support three levels of headings, please do not create a heading level below \subsubsection.
\section*{Results}
While its central ideas are unchanged, IIT updated measures of integrated information. The original formulation, IIT 1.0 \cite{Tononi2004}, underwent major developments leading to IIT 2.0 \cite{Balduzzi2008} and the latest version IIT 3.0 \cite{Oizumi2014}. In the present study, we focus on the version in IIT 2.0 \cite{Tononi2008, Balduzzi2008}, because the measure of integrated information proposed in IIT 2.0 is simpler and more feasible to calculate compared with that in IIT 3.0 \cite{Tononi2012, Oizumi2014}.

Here, we briefly review the original measure of integrated information, $\Phi$, in IIT 2.0 \cite{Tononi2008, Balduzzi2008} and describe its limitations for practical application \cite{Barrett2011}. From the concept of the original measure, we point out the lower and upper bounds that a measure of integrated information should satisfy. We introduce next two practical measures of integrated information, $\Phi_I$ and $\Phi_H$, proposed by \cite{Barrett2011} and show that $\Phi_I$ and $\Phi_H$ fail to satisfy the lower and upper bounds of integrated information. Finally, we derive a novel measure of integrated information, $\Phi^*$, from the decoding perspective, which is properly bounded from below and above.

\subsection*{Measure of integrated information with the maximum entropy distribution}
Integrated information is a quantity that measures how much extra information is generated by the system as a whole above and beyond the information independently generated by its parts \cite{Tononi2008,Balduzzi2008}. Consider partitioning a system into $m$ parts such as $M_1$, $M_2$, $\cdots$, and $M_m$ and computing the quantity of information that is integrated across the $m$ parts of a system. As detailed in Methods, the measure of integrated information proposed in IIT 2.0 can be expressed as follows:
\begin{equation}
\Phi = I(^{\max} X^{t-\tau};X^t)- \sum_{i=1}^m I(^{\max} M_i^{t-\tau}; M_i^t), \label{eq:original1}
\end{equation}
where $X^{t-\tau}$ and $X^t$ are states of a system in the past $t-\tau$ ($\tau>0$) and present $t$, respectively. The distribution of past states is assumed as the maximum entropy distribution, and the upper subscript $^{\max}$ is placed left of $X^{t-\tau}$ to explicitly indicate that the distribution of past states represents the maximum entropy distribution. The first term of Eq. \ref{eq:original1}, $I(^{\max} X^{t-\tau};X^t)$, represents the mutual information between the past and present states in the whole system, and the second term represents the sum of the mutual information between the past and present states in the $i$-th part of the system $I(^{\max} M_i^{t-\tau}; M_i^t)$. Thus, $\Phi$, the difference between them, gives the information generated by the whole system above and beyond the information generated independently by its parts. If the parts are independent, no extra information is generated, and the integrated information is $0$. We can rewrite Eq. \ref{eq:original1} in terms of entropy $H$ as follows:
\begin{equation}
\Phi = \sum_{i=1}^m H(^{\max} M_i^{t-\tau}|M_i^t) -H(^{\max} X^{t-\tau}|X^t).\label{eq:original2}
\end{equation}
To derive the above expression, we use the fact that the entropy of the whole system $H(^{\max} X^{t-\tau})$ equals the sum of the entropy of the subsystems $\sum_{i=1}^m H(^{\max} M_i^{t-\tau})$ when the maximum entropy distribution is assumed.

\subsection*{Theoretical requirements as a measure of integrated information}
To interpret a measure of integrated information as the ``extra'' information generated by a system as a whole above and beyond its parts, it should satisfy the theoretical requirements, as follows: first, integrated information should not be negative because information independently generated by the parts should never exceed information generated by the whole. Integrated information should equal 0 when the amount of information generated by the whole system equals 0 (no information) or when the amount of information generated by the whole is equal to that generated by its parts (no integration). Second, integrated information should not exceed the amount of information generated by the whole system because the information generated by the parts should be larger than or equal to 0. In short, integrated information should be lower-bounded by 0 and upper-bounded by the information generated by the whole system.

One can check the original measure $\Phi$ satisfies the lower and upper bounds.
\begin{equation}
0 \leq \Phi \leq I(^{\max} X^{t-\tau};X^t). \label{eq:}
\end{equation}
As shown in Methods, $\Phi$ can be written as the Kullback-Leibler divergence (see Eq. \ref{eq:phi}). Thus, $\Phi$ is positive or equal to 0. Further, as can be seen from Eq. \ref{eq:original1}, the upper bound of $\Phi$ is the mutual information in the entire system, because the sum of mutual information in the parts is larger than or equal to 0.

\subsubsection*{Practical measures of integrated information with empirical distribution}
As we described in the previous section, in the original measure $\Phi$, the distribution of past states is assumed as the maximum entropy distribution, which limits the practical application of $\Phi$. First, the maximum entropy distribution can be applied only when the states of a system are discrete. If the states are represented by discrete variables, the maximum entropy distribution is the uniform distribution over all possible states of $X^{t-\tau}$. When the states of a system are described by continuous variables, the maximum entropy distribution cannot be uniquely defined \cite{Barrett2011,Cover1991}. Second, the transition probability matrix of a system, $p(X^t|X^{t-\tau})$ must be known for all possible past states $X^{t-\tau}$, because the sum of $\log p(X^t|X^{t-\tau})$ over all possible past states must be computed for computing the mutual information $I(^{\max} X^{t-\tau};X^t)$. However, it is nearly impossible to estimate experimentally such a complete transition probability matrix in an actual neural system, because some states may not occur during a reasonable period of observation. Although it may be possible to force the system into a particular state by stimulating some neurons while silencing others and estimating transition probabilities for each state, this is technically extremely demanding.

A simple remedy for the limitations of the original measure $\Phi$ is to not impose the maximum entropy distribution on past states but to instead use the probability distributions obtained from empirical observations of the system. Barrett and Seth \cite{Barrett2011} adopted this strategy to derive two practical measures of integrated information from Eqs. \ref{eq:original1} and \ref{eq:original2} by substituting the maximum entropy distribution with the empirical distribution as follows:
\begin{equation}
\Phi_I= I(X^{t-\tau};X^t)- \sum_{i=1}^m I(M_i^{t-\tau}; M_i^t), \label{eq:phiI}
\end{equation}
\begin{equation}
\Phi_H = \sum_{i=1}^m H(M_i^{t-\tau}|M_i^t) -H(X^{t-\tau}|X^t). \label{eq:phiH}
\end{equation}
Note that $\Phi_I$ and $\Phi_H$ are not equal when the empirical distribution is used for past states, because the entropy of the whole system $H(X^{t-\tau})$ is not equal to the sum of the entropy of the subsystems, $\sum_i H(M_i^{t-\tau})$. $\Phi_H$ was also derived from a different perspective from IIT, i.e. the perspective of information geometry, as a measure of spatio-temporal interdependencies and is termed ``stochastic interaction'' \cite{Ay2001}.

Although these two measures appear as natural modifications of the original measure, they do not satisfy the theoretical requirements as a measure of integrated information. We discuss the problems of $\Phi_I$ and $\Phi_H$ in detail below.

\subsubsection*{Integrated information measure based on mismatched decoding}
\begin{figure}[t!]
\begin{center}
\centerline{\includegraphics[width=0.6\textwidth]{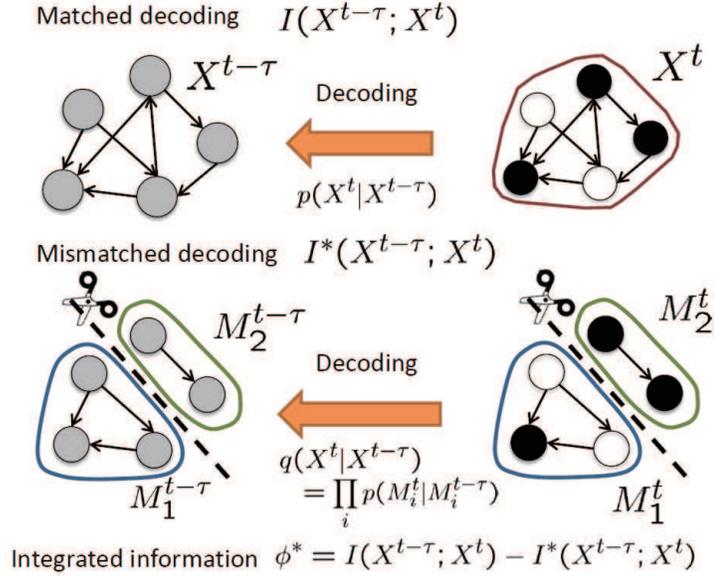}}
\caption{Integrated information with an empirical distribution based on the concept of mismatched decoding. The figure shows a system with five neurons in which the arrows represent directed connectivity and the colors represent the states of the neurons (black: silence, white: firing, gray: unknown). The past states $X^{t-\tau}$ are decoded given the present states $X^{t}$. The ``true'' conditional distribution $p(X^{t}|X^{t-\tau})$ is used for matched decoding, while a ``false'' conditional distribution $q(X^{t}|X^{t-\tau})$ is used for mismatched decoding where the parts of a system $M_1$ and $M_2$ are assumed independent. The amount of information about past states that can be extracted from present states using matched and mismatched decoding is quantified by the mutual information $I(X^{t-\tau};X^t)$ and the ``hypothetical'' mutual information $I^*(X^{t-\tau};X^t)$ for mismatched decoding, respectively. In this framework, integrated information, $\Phi^*(X^{t-\tau};X^t)$, is defined as the difference between $I(X^{t-\tau};X^t)$ and $I^*(X^{t-\tau};X^t)$. } \label{fig:mismatch}
\end{center}
\end{figure}

Here, we propose an alternative practical measure of integrated information that satisfies the theoretical requirements which we call $\Phi^*$ (phi star) (Fig. \ref{fig:mismatch}). $\Phi^*$, which uses the empirical distribution, can be applied to actual neuronal recordings. Similar to $\Phi_I$, we will derive $\Phi^*$ based on the original measure $\Phi$ in Eq. \ref{eq:original1} based on mutual information. Given the problem of $\Phi_I$ in Eq. \ref{eq:phiI}, we should refine the second term of Eq. \ref{eq:phiI}, while the first term, the mutual information in the whole system, is unchanged. The second term should be a quantity that can be interpreted as information generated independently by the parts of a system and should be less than information generated by the system as a whole.

To derive a proper second term in Eq. \ref{eq:phiI}, we interpret the mutual information from a decoding perspective and introduce the concept of ``mismatched decoding'', which was developed by information theory \cite{Merhav1994} (see Methods for details). Consider that the past states $X^{t-\tau}$ are decoded given the present states $X^t$. From the decoding perspective, the mutual information can be interpreted as the maximum information about the past states that can be obtained knowing the present states. To extract the maximum information, the decoding must be performed optimally using the ``true'' conditional distribution,
\begin{equation}
p(X^t|X^{t-\tau} )=p(M_1^t,\cdots,M_m^t|M_1^{t-\tau},\cdots,M_m^{t-\tau}).
\end{equation}
Note that the expression on the right accounts explicitly for interactions among all the parts. The optimal decoding can be performed using maximum likelihood estimation. In the above setting, the maximum likelihood estimation means choosing the past state that maximizes $p(X^t|X^{t-\tau})$ given a present state. Decoding that uses the true distribution, $p(X^t|X^{t-\tau})$, is called ``matched decoding'' because the probability distribution used for decoding matches the actual probability distribution.

Decoding that uses a ``false'' conditional distribution, $q(X^t|X^{t-\tau})$, is called ``mismatched'' decoding. To quantify integrated information, we consider specifically the mismatched decoding that uses the ``partitioned'' probability distribution $q(X^t|X^{t-\tau})$,
\begin{equation}
q(X^t|X^{t-\tau})= \prod_{i=1}^m p(M_i^t|M_i^{t-\tau}), \label{eq:}
\end{equation}
where a system is partitioned into parts and the parts $M_i$ are assumed as independent. $q(X^t|X^{t-\tau})$ is the product of the conditional probability distribution in each part $p(M_i^t|M_i^{t-\tau})$. The distribution, $q(X^t|X^{t-\tau})$, is ``mismatched'' with the actual probability distribution, because parts are generally not independent. We evaluate the amount of information obtained from mismatched decoding. As is matched decoding, mismatched decoding is also performed using the maximum likelihood estimation, wherein the past state that maximizes $q(X^t|X^{t-\tau})$ is selected. The amount of information obtained from mismatched decoding is necessarily degraded compared with that obtained from matched decoding. The best decoding performance can be achieved only using matched decoding with the actual probability distribution $p(X^t|X^{t-\tau})$.

We consider the amount of information that can be obtained from mismatched decoding, $I^* (X^{t-\tau};X^t)$, as a proper second term of Eq. \ref{eq:phiI} (see Methods for the mathematical expression of $I^*$). The difference between $I(X^{t-\tau};X^t)$ and $I^*(X^{t-\tau};X^t)$ provides a new practical measure of integrated information (Fig. \ref{fig:mismatch}),
\begin{equation}
\Phi^*(X^{t-\tau};X^t) = I(X^{t-\tau};X^t) - I^*(X^{t-\tau};X^t). \label{eq:phistar}
\end{equation}
$\Phi^*$ quantifies the information loss caused by mismatched decoding where a system is partitioned into independent parts, and the interactions between the parts are ignored. $\Phi^*$ satisfies the theoretical requirements as a measure of integrated information, because $I^*$ is greater than or equal to 0 and is less than or equal to the information in the whole system $I$. $\Phi^*$ defined this way is equivalent to the original measure $\Phi$ if the maximum entropy distribution is imposed on past states instead of an empirical distribution (see Supporting Information for the proof). Thus, we can consider $\Phi^*$ as a natural extension of the original measure $\Phi$ to the case when the empirical distribution is used.

\subsection*{Analytical computation of $\Phi^*$ using Gaussian approximation}
Although using an empirical distribution instead of the maximum entropy distribution makes integrated information more feasible to calculate, it is still difficult to compute $\Phi^*$ in a large system, because the summation (or integral) over all possible states must be calculated. The number of all possible states grows exponentially with the size of the system and therefore, computational costs for computing $\Phi^*$ also grow exponentially. Thus, for practical calculation of $\Phi^*$, we need to approximate $\Phi^*$ in some way such as approximating the probability distribution of neural states using the Gaussian distribution \cite{Barrett2011}. $\Phi^*$ can be analytically computed using the Gaussian approximation (see Methods). The Gaussian approximation reduces significantly the computational costs and makes $\Phi^*$ practically computable even in a large system. 

\subsection*{Theoretical requirements are not satisfied by previously proposed measures}
As described above, the lower and upper bounds of integrated information should equal 0 and the information generated by the whole system, respectively. In this section, by considering two extreme cases, we demonstrate that the previously proposed measures $\Phi_H$ and $\Phi_I$ \cite{Barrett2011} do not satisfy either the lower or upper bound.

\subsubsection*{When there is no information}
First, we consider cases where there is no information between past and present states of a system, i.e. $I(X^{t-\tau};X^t)=0$. In this case, integrated information should be 0. As expected, $\Phi^*$ and $\Phi_I$ are 0, because the amount of information for mismatched decoding, $I^*(X^{t-\tau};X^t)$, and the mutual information in each part, $I(M_i^{t-\tau};M_i^t)$ are both 0 when $I(X^{t-\tau};X^t)=0$.
\begin{align}
\Phi^* &= 0, \\
\Phi_I &= 0.
\end{align}
However, $\Phi_H$ is not 0. $\Phi_H$ can be written as
\begin{equation}
\Phi_H = \sum_i H(M^{t-\tau}_i) - H(X^{t-\tau}). \label{eq:noinf}
\end{equation}
$\Phi_H$ is not 0 when the information is 0 because $\Phi_H$ is not based on the mutual information but on the conditional entropy (see Eq. \ref{eq:phiH}). Therefore, $\Phi_H$ does not necessarily reflect the amount of information in a system.

As a simple example that shows the above problem of $\Phi_H$, consider the  following linear regression model, 
\begin{equation}
X^t = A \cdot X^{t-1} + E^t. \label{eq:AR}
\end{equation}
Here, $X$ is the state of units, $A$ is a connectivity matrix, and $E^t$ is multivariate Gaussian noise with zero mean and covariance $\Sigma(E)$. $E^t$ is uncorrelated over time. For simplicity, consider a system composed of two units (the following argument can be easily generalized to a system with more than two units). We set the connectivity matrix, $A$, and the covariance matrix of noise, $\Sigma(E)$ as follows:
\begin{equation}
A = a \cdot
\left(
\begin{array}{cc}
1 & 1 \\
1 & 1
\end{array}
\right). \label{eq:A}
\end{equation}
\begin{equation}
\Sigma(E) = 
\left(
\begin{array}{cc}
1 & c \\
c & 1
\end{array}
\right),
\end{equation}
where $a$ and $c$ are parameters that control the strengths of connections and noise correlation, respectively. We compute measures of integrated information using the above model. The time difference $\tau$ is set to 1. We assume that the prior distribution of the system is the steady state distribution, where the covariance of past states, $\Sigma(X^{t-1})$, and that of present states, $\Sigma(X^{t})$, are equal, i.e. $\Sigma(X^{t-1})= \Sigma(X^{t})= \Sigma(X)$. The covariance of the steady state distribution $\Sigma(X)$ can be calculated by taking the covariance of both sides of Eq. \ref{eq:AR},
\begin{equation}
\Sigma(X) = A \Sigma(X) A^T + \Sigma(E).
\end{equation}

\begin{figure}[t!]
\begin{center}
\centerline{\includegraphics[width=0.6\textwidth]{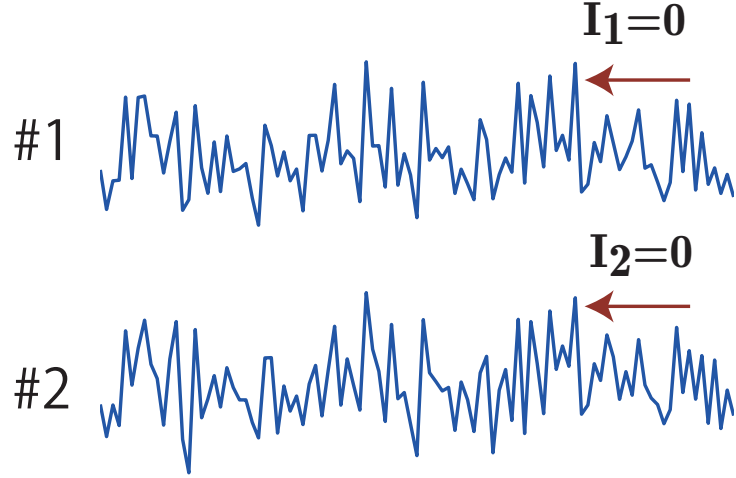}}
\caption{Exemplar time series when the strength of noise correlation c and the connection strength a are set to 0.9 and 0, respectively in the linear regression model (Eq. \ref{eq:AR}). $I_1$ and $I_2$ represent the mutual information in units 1 and 2. Because there is no connection, there is no information between past and present states of the system: $I_1$ and $I_2$ are both 0. In this case, $\Phi^*$ and $\Phi_I$ are 0 as they should be, yet $\Phi_H$ is positive.} \label{fig:time_series_noinfo}
\end{center}
\end{figure}

\begin{figure}[t!]
\begin{center}
\centerline{\includegraphics[width=0.8\textwidth]{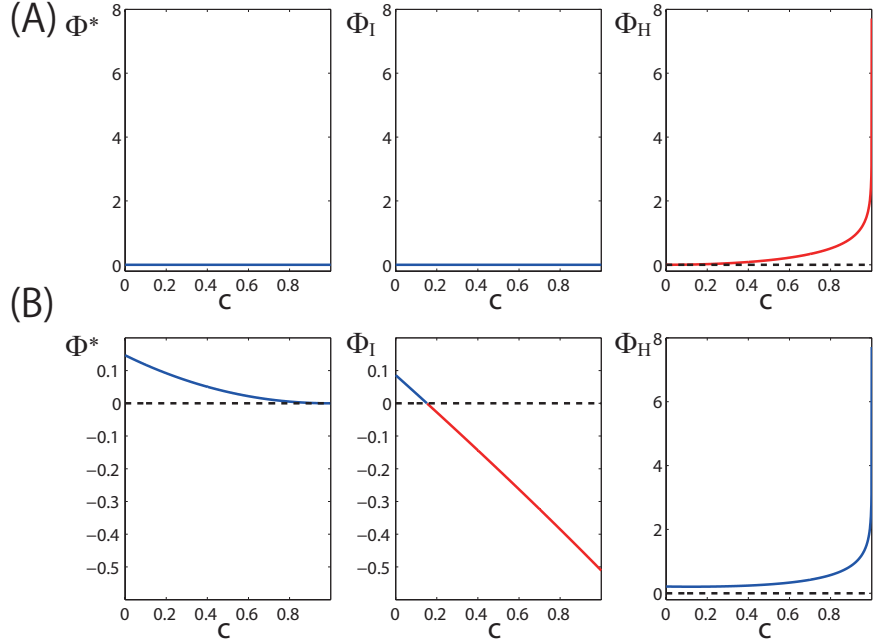}}
\caption{Theoretical requirements as a measure of integrated information are not satisfied by $\Phi_H$ and $\Phi_I$. The behaviors of $\Phi^*$, $\Phi_I$, and $\Phi_H$ are shown in the left, middle, and right panels, respectively, when the strength of noise correlation $c$ is varied in a linear regression model (Eq. \ref{eq:AR}). Red lines indicate the regime where the theoretical requirement is violated, and the blue lines indicate that the theoretical requirement is satisfied. Dotted black lines are drawn at 0. (A) Violation of the upper bound. The strength of connections $a$ is set to 0. In this case, there is no information between past and present states of the system but $\Phi_H$ is not 0, i.e., $\Phi_H$ violates the upper bound. (B) Violation of the lower bound. The strength of connections $a$ is set to 0.4. At the right ends of the figures where $c$ is 1, the two units in the system are perfectly correlated. $\Phi_I$ is negative, i.e., violates the lower bound when the degree of correlation is high.} \label{fig:phi_comp}
\end{center}
\end{figure}

We consider a case where the connection strength $a$ is 0. Fig. \ref{fig:time_series_noinfo} shows an exemplar time series when the strength of noise correlation $c$ is 0.9. Because there are no connections, including self-connections within each unit, each unit has no information between past and present states, i.e., $I_1=I_2=0$. As can be seen from Fig. \ref{fig:time_series_noinfo}, however, the two time series correlate at each moment because of the high noise correlation.

We varied the degree of noise correlation, $c$, from 0 to 1 while keeping the connection strength $a$ as 0 (Fig. \ref{fig:phi_comp}(A)). $\Phi^*$ and $\Phi_I$ stay 0 independent of noise correlation. However, an entropy-based measure, $\Phi_H$, increases monotonically with $c$, irrespective of the amount of information in the whole system (Fig. \ref{fig:phi_comp}(A)). In other words, $\Phi_H$ does not reflect the amount of information in a system, but does reflect the degree of correlation between the parts. As shown in Eq. \ref{eq:noinf}, $\Phi_H$ is the difference between the sum of entropy within each part and entropy in the whole system. When the parts correlate, the entropy in the whole system decreases. In contrast, the sum of entropy of each part does not change, because the degree of noise within each part (the diagonal elements of $E^t$) is fixed. Thus, $\Phi_H$ increases as the degree of noise correlation $c$ increases.

Because $\Phi_H$ is not 0 even when there is no information in the system, we can see that it exceeds the mutual information in the whole system and does not satisfy the upper bound as a measure of integrated information.

\subsubsection*{When parts are perfectly correlated}
\begin{figure}[t!]
\begin{center}
\centerline{\includegraphics[width=0.6\textwidth]{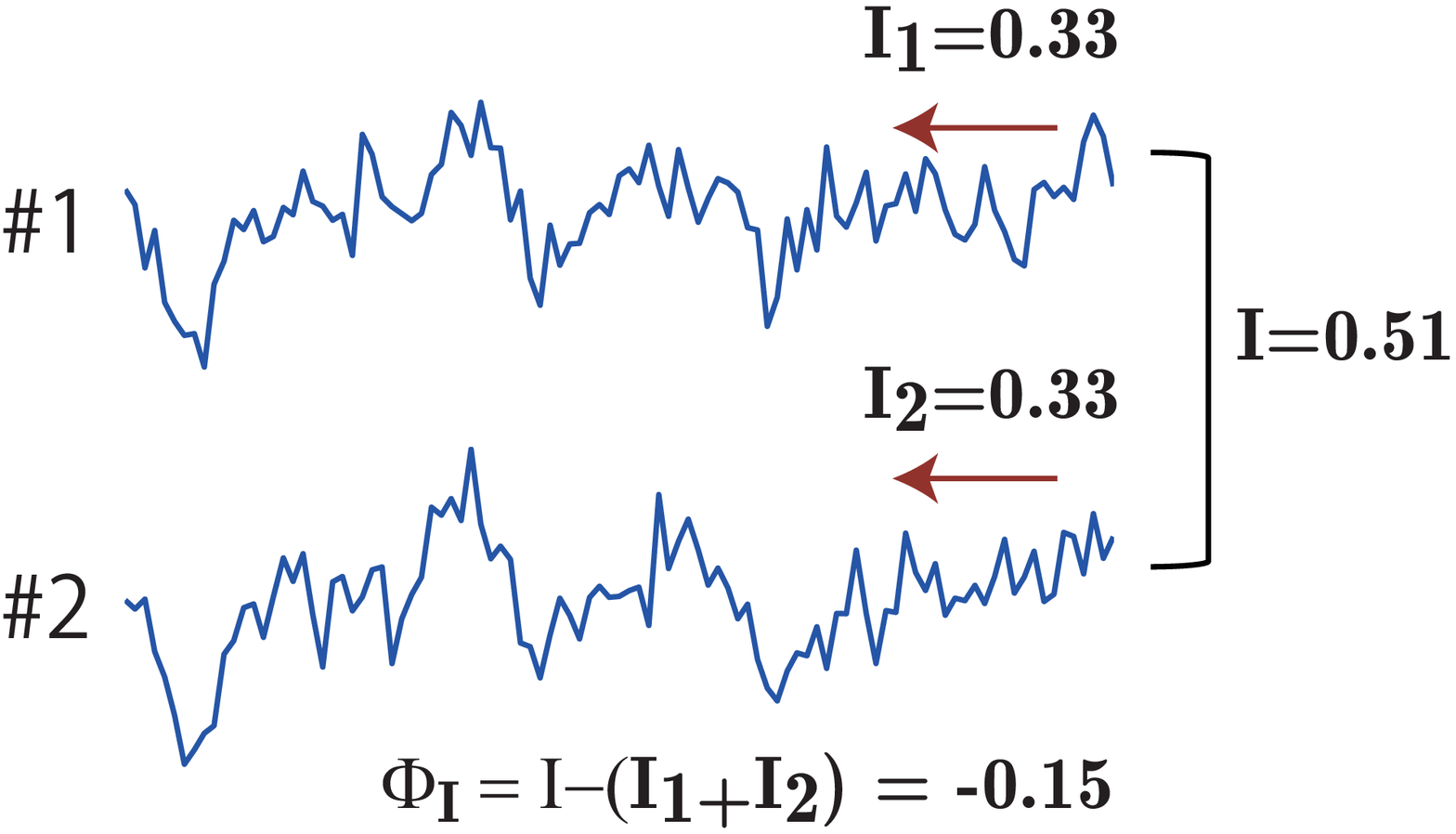}}
\caption{Exemplar time series when the strength of noise correlation $c$ and the connection strength $a$ are set to both 0.4 in the linear regression model (Eq. \ref{eq:AR}). $I_1$ and $I_2$ represent the mutual information in units 1 and 2, and $I$ represents the mutual information in the whole system. In this case, the sum of the mutual information in the parts exceeds the mutual information in the whole system and $\Phi_I$ is negative. } \label{fig:time_series_corr}
\end{center}
\end{figure}

Next, we consider a case where the parts are perfectly correlated. More specifically, consider the case where the two parts $M_1$ and $M_2$ are equal at every time, i.e. $M^{t-\tau}_1=M^{t-\tau}_2=M^{t-\tau}$ and $M^{t}_1=M^{t}_2=M^t$. Here, $\Phi^*$ is 0 because the amount of information extracted by mismatched decoding would not degrade even if the other part is ignored for decoding (see Supporting Information for the mathematical proof).
\begin{equation}
\Phi^* = 0. 
\end{equation}
Regarding $\Phi_I$, when the parts are perfectly correlated, the mutual information of each part is equal to each other, $I(M_1^{t-\tau};M_1^t)=I(M_2^{t-\tau};M_2^t)=I(M^{t-\tau};M^t)$ and the mutual information in the whole system is equal to the mutual information of each part, $I(X^{t-\tau};X^t)=I(M^{t-\tau};M^t)$. Thus, the second term in Eq. \ref{eq:phiI} is twice the value of the first, and $\Phi_I$ is the negative value of the mutual information in one part,
\begin{equation}
\Phi_I = - I(M^{t-\tau};M^t).
\end{equation}
Thus, $\Phi_I$ does not satisfy the lower bound as a measure of integrated information. $\Phi_H$ is given by
\begin{equation}
\Phi_H = H(X^{t-\tau}|X^t) - 2 H(M^{t-\tau}|M^t),
\end{equation}
which is larger than or equal to 0 ($\Phi_H$ is always larger than or equal to 0 because it can be written as the Kullback-Leibler divergence.).

To illustrate the behaviors of these three measures of integrated information when the degree of correlation varies, we considered the same linear regression model presented in the previous section (Eq. \ref{eq:AR}). We varied the degree of noise correlation, $c$, from 0 to 1 while keeping connection strength $a$ as 0.4. When $c$ is 1, the two units correlate perfectly. Fig. \ref{fig:time_series_corr} shows an exemplar time series when $c$ is 0.4 and $a$ is 0.4. $\Phi_I$ takes positive values when $c$ is less than $\sim0.2$ but takes negative values when $c$ is greater (Fig.\ref{fig:phi_comp}(B)). $\Phi^*$ decreases monotonically with $c$ and becomes 0 when $c$ is 1. $\Phi_H$ increases monotonically with $c$ reflecting the degree of correlation between the units. The detailed behaviors of $\Phi^*$, $\Phi_I$ and $\Phi_H$ when $a$ and $c$ are both varied are shown in Supporting Information.

\section*{Discussion}
In this study, we consider two theoretical requirements that a measure of integrated information should satisfy, as follows: the lower and upper bounds of integrated information should be 0 and the amount of information generated by the whole system, respectively. The theoretical requirements are naturally derived from the original philosophy of integrated information \cite{Tononi2008, Balduzzi2008}, which states that integrated information is the information generated by a system as a whole above and beyond its parts. The original measure of integrated information $\Phi$ satisfies the theoretical requirements that are required so that we can interpret a measure of integrated information according to the original philosophy. To derive a practical measure of integrated information that satisfies the required lower and upper bounds, we introduced a concept of mismatched decoding. We defined our measure of integrated information $\Phi^*$ as the amount of information lost when a mismatched probability distribution, where a system is partitioned into ``independent'' parts, is used for decoding instead of the actual probability distribution. In this framework, $\Phi^*$ quantifies the amount of information loss associated with mismatched decoding where all interactions between the parts of a system are ignored and therefore quantifies the amount of information integrated by such interactions between the parts. We show that $\Phi^*$ satisfies the lower and upper bounds, that $\Phi_I$ does not satisfy the lower bound, and that $\Phi_H$ does not satisfy the upper bound. We consider $\Phi^*$ a proper measure of integrated information that can be generally used for practical applications.

The basic concept of Integrated Information Theory (IIT) was tested by conducting empirical experiments, and the evidence accumulated supports the conclusion that when consciousness is lost, integration of information is lost \cite{Massimini2005, Massimini2007, Ferrarelli2010, Rosanova2012, Casali2013}. In particular, Casali and colleagues \cite{Casali2013} found that a complexity measure, motivated by IIT, successfully separates conscious awake states from various unconscious states due to deep sleep, anesthesia, and traumatic brain injuries. Although their measure is inspired by the concept of integrated information, it measures the complexity of averaged neural responses to one particular type of external perturbation (e.g. a TMS pulse to a target region) and does not directly measure integrated information.

There are few studies that directly estimate integrated information in the brain \cite{Lee2009,Chang2012} using the measure introduced in IIT 1.0 \cite{Tononi2004} or $\Phi_H$. Our new measure of integrated information, $\Phi^*$, will contribute to experiments designed to test whether integrated information is a key to distinguishing conscious states from unconscious states \cite{Alkire2008,Boly2011,Sanders2012}. 

We considered the measure of integrated information proposed in IIT 2.0 \cite{Tononi2008,Balduzzi2008}, because its computations are feasible. There are several updates in the latest version, IIT 3.0 \cite{Oizumi2014}. One important update is that both the cause and effect of a present state are considered for quantifying integrated information. In IIT 2.0, integrated information is quantified by measuring how the distribution of past states differs when a present state is given, i.e. only the cause of a present state is considered (see Methods). Moreover, IIT 3.0 measures how the distribution of future states differs when a present state is given, i.e. the effect of a present state is considered. Our measure $\Phi^*$ does not asymmetrically treat the past cause and the future effect when a present state is given, because the mutual information is a symmetric measure for the times $t-\tau$ and $t$. An unanswered question is how integrated information should be practically calculated taking cause and effect into account separately, using an empirical distribution.

An unresolved difficulty that impedes practical calculation of integrated information is how to partition a system. In the present study, we considered only the quantification of integrated information when a partition of a system is given. IIT requires that integrated information should be quantified using the partition where information is least integrated, called the minimum information partition (MIP) \cite{Tononi2008,Balduzzi2008}. To find the MIP, every possible partition must be examined, yet the number of possible partitions grows exponentially with the size of the system.  One way to work around this difficulty would be to develop optimization algorithms to quickly find a partition that well approximates the MIP.

Besides the practical problem of finding the MIP, there remains a theoretical problem of how to compare integrated information across different partitions. Integrated information increases as the number of parts gets larger, because more information will be lost by partitioning the system. Further, integrated information is expected to be larger in a symmetric partition where a system is partitioned into two parts of equal size than in an asymmetric partition. IIT 2.0 \cite{Balduzzi2008} proposes a normalization factor, which considers these issues. However, there might be other possible ways to perform normalization. It is unclear whether there is a reasonable theoretical foundation that adjudicates the best normalization scheme. Moreover, it is unclear if the normalization factor, which was proposed under the assumption that the states of a system are represented by discrete variables, would be appropriate for the cases where the states are represented by continuous variables. Further investigations are required to resolve practical and theoretical issues related to the MIP.

Although we derived $\Phi^*$, because we were motivated by IIT and its potential relevance to consciousness, $\Phi^*$ has unique meaning from the perspective of information theory, which is independent of IIT. Thus, it can be applied to research fields other than research on consciousness. $\Phi^*$ quantifies the loss of information when interactions or connections between the units in a system are ignored. Thus, $\Phi^*$ can be expected to be related to connectivity measures such as Granger causality \cite{Ding2006} or transfer entropy \cite{Vicente2011}. It will be interesting to clarify mathematical relationships between $\Phi^*$ and the other connectivity measures. Here, we indicate only an apparent difference between them as follows: $\Phi^*$ intends to measure global integrations in a system as a whole, while traditional bivariate measures such as Granger causality or transfer entropy intends to measure local interactions between elements of the system. Consider that we divide a system into parts $A$, $B$, and $C$. Using integrated information, our goal is to quantify the information integrated among $A$, $B$, and $C$ as a whole. In contrast, what we quantify using Granger causality or transfer entropy analysis is the influence of $A$ on $B$, $B$ on $C$, $C$ on $A$ and the reverse. It is not obvious how a measure of global interactions in the whole system should be defined and derived theoretically from measures of local interaction. As an example, one possibility is simply summing up all of local interactions and considering the sum as a global measure \cite{Seth2011}. Yet, more research is required to determine whether such an approach is a valid method to define global interactions. $\Phi^*$, in contrast, is not derived from local interaction measures but is derived directly by comparing the total mutual information in the whole system with hypothetical mutual information when the system is assumed to be partitioned into independent parts. Thus, the interpretation of $\Phi^*$ is straightforward from an information theoretical viewpoint. Our measure, which we consider a measure of global interactions, may provide new insights into diverse research subjects as a novel tool for network analysis.

% You may title this section "Methods" or "Models".
% "Models" is not a valid title for PLoS ONE authors. However, PLoS ONE
% authors may use "Analysis"
\section*{Methods}
\subsubsection*{Intrinsic and extrinsic information}
Before introducing the concept of integrated information, we clarify the definition of ``information'' in IIT. In IIT, information always refers to intrinsic information in contrast to extrinsic information \cite{Oizumi2014}. Intrinsic and extrinsic here refers to the perspective from which information is considered. Intrinsic information is quantified from the perspective of the system itself while extrinsic information is quantified from the perspective of an external observer. In this section, we explain the differences in detail.

In neuroscience, many researches focus on quantifying the informational relationship between neural states and external stimuli or observable output behaviors \cite{Rieke1997,Dayan2001,Averbeck2006,Quiroga2009}. For example, the mutual information between neural states $X$ and external stimuli $S$ is quantified as
\begin{equation}
I(X;S) = H(S) - H(S|X). \label{eq:extrinsic}
\end{equation}
where the entropy $H(S)$ and the conditional entropy $H(S|X)$ are given by
\begin{align}
H(S) &= -\sum_s p(s) \log p(s), \\
H(S|X) &= -\sum_{x,s} p(s,x) \log p(s|x).
\end{align}
Here, $x$ and $s$ represent a particular neural state and a particular external stimulus, respectively, with $p(x)$, $p(s)$, $p(s,x)$, and $p(s|x)$ denoting the probability of $x$ and $s$, the joint probability of $x$ and $s$, and a conditional probability of $s$ given $x$. The sum is calculated for all possible neural states $x$ or over all stimuli $s$. The capital $S$ and $X$ represent an entire set of $s$ or $x$, respectively. When we assume that continuous variables represent neural states, we must replace the sum $\sum$ with the integral $\int$. As shown in Eq. \ref{eq:extrinsic}, mutual information is expressed as the difference between the entropy of stimuli, $H(S)$, and the conditional entropy of stimuli given neural states, $H(S|X)$. Thus, $I(X;S)$ quantifies the reduction of uncertainty about stimuli by acquiring knowledge of neural states from the perspective of an external observer, i.e. to what extent can an external observer know about external stimuli by observing neural states. This type of information is called extrinsic information because the information is quantified from an external observer's point of view. 

Intrinsic information, in contrast, is quantified from the viewpoint of the system itself, independent of observations by any other external entity \cite{Oizumi2014}. Intrinsic information should not depend on external variables but only on internal variables of the system. If information concerns consciousness, it is considered intrinsic information, because consciousness is independent of external observers. With this concept of intrinsic information, IIT aims to quantify how much ``difference'' the internal mechanisms of a system makes for the system itself, i.e. the degree of influence a system exerts on itself through its internal causal mechanisms. How the past states would affect present states can be determined by the transition probability matrix of the system, $p(X^t|X^{t-\tau})$, which specifies probabilities according to which any state of a system transits to any other state. Here, $X^t$ and $X^{t-\tau}$ are states of the system at times $t$ and $t-\tau$, which we call present and past states, respectively. ITT quantifies intrinsic information using the transition probability matrix.

The intrinsic information proposed in IIT 2.0 quantifies to what extent the mechanisms of the system make the posterior probability distribution of past states given a present state different compared with a prior distribution of past states. The posterior probability distribution of past states given a present state represents the likelihood of potential causes of the given present state. Intrinsic information in IIT 2.0, which is called ``effective information'', is defined as the difference between the posterior probability distribution, $p(X^{t-\tau}|x^t)$, and a prior distribution of past states, $p(X^{t-\tau})$ as follows:
\begin{equation}
ei(x^t) = D_{KL} \left( p(X^{t-\tau}|x^t) || p(X^{t-\tau}) \right), \label{eq:ei}
\end{equation}
where $D_{KL}(p(X)||q(X))$ is the Kullback-Leibler divergence, which measures the distance between the two probability distributions $p$ and $q$ and is given by
\begin{equation}
D_{KL}(p(X)||q(X)) = \sum_x p(x) \log \frac{p(x)}{q(x)}. \label{eq:KL}
\end{equation}
If there are no causal mechanisms within the system, present states are not affected by past states. Thus, the posterior distribution of past states does not differ from the prior distribution. IIT interprets the degree of the ``difference'' made in the posterior probability distribution of past states according to its internal mechanisms, as information generated intrinsically within the system. Note that while intrinsic information is based on an intrinsic property of the system, it does not mean that it cannot be quantified by an external observer.

To quantify intrinsic information, in addition to the transition probability matrix, a prior distribution of past states must be specified. Although the transition probability matrix is determined by the intrinsic mechanisms of a system, a prior distribution of past states cannot be uniquely determined. There are many possible methods to choose a prior distribution from different standards. For example, in the context of channel capacity in information theory, the prior distribution that maximizes information may be selected \cite{Cover1991}. In contrast, IIT selects the maximum entropy distribution as a prior distribution \cite{Jaynes1957, Balduzzi2008}. If a system's states are represented as a set of discrete variables, the maximum entropy distribution is the uniform distribution over all possible past states $X^{t-\tau}$. Thus, using the maximum entropy distribution as a prior distribution means that every possible past state is equally likely as a cause of a present state.

Although the maximum entropy distribution can be uniquely defined for discrete variables, this is not possible for continuous variables \cite{Cover1991,Barrett2011}. If some constraints are given, the maximum entropy distribution can be defined for continuous variables. For example, under the constraints that the mean and the variance of the variables are fixed at specific values, the Gaussian distribution with the specified mean and variance is the maximum entropy distribution. There is no principle that determines what types of constraints should be imposed and how the maximum entropy distribution should be uniquely determined for continuous variables. Thus, intrinsic information (and integrated information) defined in IIT 2.0 can be applied only to discrete variables.

Using entropy, Eq. \ref{eq:ei} can be written as
\begin{equation}
ei(x^t) = H( p(^{\max}X^{t-\tau}) ) - H(p(^{\max}X^{t-\tau}|x^t)), \label{eq:ei_H}
\end{equation}
where the upper subscript $\max$ placed on the left side of $X^{t-\tau}$ is a reminder that the distribution of $X^{t-\tau}$ is the maximum entropy distribution. Eq. \ref{eq:ei_H} provides another interpretation of effective information. It quantifies to what extent uncertainty of the past states $X^{t-\tau}$ (the entropy, $H(^{\max}X^{t-\tau})$) can be reduced by knowing a particular present state $x^{t}$ from the system's intrinsic point of view. Using Bayes' rule, the posterior distribution, $p(^{\max}X^{t-\tau}|x^t)$, can be calculated as
\begin{equation}
p(^{\max}X^{t-\tau}|x^t) = \frac{p(x^t|^{\max}X^{t-\tau}) p(^{\max}X^{t-\tau})}{p(x^t)}. \label{eq:posterior}
\end{equation}

Averaging $ei(x^t)$ over all possible present states $x^t$, the averaged effective information equals the mutual information between past states and present states,
\begin{align}
EI &= \sum_{x^t} p(x^t) ei(x^t), \\
&= H(p^{\max}(X^{t-\tau})) - H(p(^{\max} X^{t-\tau}|X^t)), \\
&= I(^{\max} X^{t-\tau};X^t). \label{eq:ave_ei}
\end{align}
While effective information is originally quantified in a state-dependent manner as in Eq. \ref{eq:ei} (with a particular present state, $x^t$), we consider only the averaged effective information in Eq. \ref{eq:ave_ei} (with an entire set of present states, $X^t$) following the previous study \cite{Barrett2011}.

\subsubsection*{Integrated information}

Integrated information is the quantity that measures the information generated by the system as a whole above and beyond the information generated independently by its parts \cite{Tononi2008,Balduzzi2008}. As performed when computing information, integrated information is computed between the system's past $X^{t-\tau}$ and present states $X^t$. Consider partitioning a system into $m$ parts such as $M_1$, $M_2$, $\cdots$, and $M_m$ and computing the amount of information that is integrated across $m$ parts. Quantifying integrated information is equivalent to quantifying the amount of information lost by partitioning the system. In IIT, partitioning into $m$ parts corresponds to splitting the transition probability matrix $p(X^t|X^{t-\tau})$ into the product of each transition probability matrix in the parts $p(M_i^t|M_i^{t-\tau})$. The partitioned transition probability matrix, $q(X^t|X^{t-\tau})$, can be written as
\begin{equation}
q(X^t|X^{t-\tau})= \prod_{i=1}^m p(M_i^t|M_i^{t-\tau}). \label{eq:split}
\end{equation}
Integrated information, $\phi(x^t)$, proposed in IIT 2.0 is defined as the difference between the posterior probability distribution of past states given a present state in the intact system, $p(^{\max}X^{t-\tau}|x^t)$ and that in the ``partitioned'' system, $q(^{\max}X^{t-\tau}|x^t)$ is as follows:
\begin{equation}
\phi(x^t) = D_{KL} \left( p(^{\max} X^{t-\tau}|x^t) || q(^{\max} X^{t-\tau}|x^t) \right), \label{eq:phi}
\end{equation}
where $D_{KL}$ is the Kullback-Leibler divergence defined in Eq. \ref{eq:KL}, and past states are assumed as the maximum entropy distribution. $q(^{\max} X^{t-\tau}|x^t)$ is defined as follows:
\begin{equation}
q(^{\max}X^{t-\tau}|x^t) = \frac{q(x^t|^{\max}X^{t-\tau}) q (^{\max}X^{t-\tau})}{q(x^t)}, \label{eq:q_posterior} 
\end{equation}
where $q(x^t)= \sum_{X^{t-\tau}} q(x^t|X^{t-\tau}) q(^{\max}X^{t-\tau}) $ and $q(^{\max}X^{t-\tau})$ is the maximum entropy distribution. Integrated information defined in Eq. \ref{eq:phi} quantifies the difference in the posterior probability distribution of past states given a present state, if the parts of the system are forced to be independent.

Although the original integrated information measure $\phi(x^t)$ is defined for a particular present state $x^t$, we consider only the average of $\phi(x^t)$ over all possible states as is performed for quantifying information in the previous section. The averaged integrated information $\Phi$ can be calculated as follows:
\begin{align}
\Phi &= \sum_{x^t} p(x^t) \phi(x^t), \\
&= \sum_{x^t} p(x^t) D_{KL} \left( p(^{\max} X^{t-\tau}|x^t) || q(^{\max} X^{t-\tau}|x^t) \right), \\
&= \sum_{x^t} p(x^t) \sum_{x^{t-\tau}} p(^{\max} x^{t-\tau}|x^t) \log \frac{p(^{\max} x^{t-\tau}|x^t)}{q(^{\max} x^{t-\tau}|x^t)}.
\end{align}
Using Eq \ref{eq:split} and \ref{eq:q_posterior}, we can write $\Phi$ in terms of entropy as follows:
\begin{equation}
\Phi = \sum_{i=1}^m H(^{\max} M_i^{t-\tau}|M_i^t) - H(^{\max} X^{t-\tau}|X^t). \label{eq:originalH}
\end{equation}
As shown in Eq. \ref{eq:originalH}, integrated information measures the difference between the uncertainty of past states given present states in the intact system and that in the partitioned system. The uncertainty of the partitioned system is always larger than that of the intact system and the increase in uncertainty corresponds to the loss of information caused by partitioning. We can rewrite Eq. \ref{eq:originalH} in terms of mutual information as follows:
\begin{equation}
\Phi = I(^{\max} X^{t-\tau};X^t)- \sum_{i=1}^m I(^{\max} M_i^{t-\tau}; M_i^t), \label{eq:originalI}
\end{equation}
where we use the fact that the entropy of the whole system $H(^{\max} X^{t-\tau})$ is the same as the sum of the entropy of the subsystems $\sum_{i=1}^m H(^{\max} M_i^{t-\tau})$ when the maximum entropy distribution is assumed. 

\subsubsection*{Quantitative meaning of $I$ and $I^*$ in information theory}
In this section, we briefly review the quantitative meaning of mutual information $I$ in information theory and that of its extension to mismatched decoding $I^*$, which was developed by Merhav et al. \cite{Merhav1994} (see also \cite{Cover1991, Latham2005, Oizumi2010}). Consider information transmission over a noisy channel $p(Y|X)$ where $X$ is the input and $Y$ is the output. For simplicity, assume that $X$ and $Y$ are both 0 or 1. (In the Results section, we consider the case where $X$ and $Y$ are the past and present states of a system, $X^{t-\tau}$ and $X^t$, respectively, and the states of a system are multidimensional variables but the same arguments as described below are generally applicable to such a case.) The sender transmits a sequence of $X$ with length $N$ called a code word, $c=[X_1,X_2,\cdots, X_N]$, over the noisy channel. For binary inputs, there are $2^N$ possible code words, but the sender does not transmit them all. A set of the code words transmitted over the noisy channel is called a codebook. The codebook is shared between the sender and the receiver. The transmitted code word is disturbed by the noise that depends on $p(Y|X)$ and is changed to $c'=[Y_1,Y_2,\cdots, Y_N]$, where $Y_i$ is the output of $X_i$. The job of the receiver is to infer (decode) which code word is sent from the received message $c'$. Consider the question as follows: For the receiver to decode the message ``error-free'' (more precisely, with an infinitesimally small error with limits of $N \to \infty$), how many code words can the sender transmit, or how many code words can the codebook contain?

Shannon's noisy channel coding theorem answers this question. According to the noisy channel coding theorem, the mutual information determines the upper limit of the number of code words that can be sent error-free over a noisy channel. We denote the maximal number of code words that can be sent error-free over the noisy channel by $2^{RN}$, where $R$ is called the information transfer rate and is less than or equal to 1. The information transfer rate $R$ is given by the mutual information $I$ between $X$ and $Y$,
\begin{equation}
R = I(X;Y).
\end{equation}
To achieve the maximal information transfer rate given by the mutual information, the receiver must optimally decode a message, which can be performed using the maximum likelihood estimation. The maximum likelihood estimation means choosing the code word $c$ in the codebook that maximizes the likelihood $p(c'|c)$,
\begin{equation}
p(c'|c) = \prod_i p(Y_i|X_i).
\end{equation}
Note that the optimal decoding scheme uses the actual probability distribution $p(Y|X)$. This type of decoding is called matched decoding, because the probability distribution used for decoding is matched with the actual probability distribution. If a mismatched probability distribution $q(Y|X)$, which is different from the actual probability distribution $p(Y|X)$, is used for decoding instead, the information transfer rate necessarily degrades. The information transfer rate $R^*$ for a mismatched decoding is given by $I^*$,
\begin{equation}
R^* = I^*(X;Y).
\end{equation}
As in matched decoding, decoding is performed using the maximum likelihood estimation with the following ``mismatched'' likelihood function $q(c'|c)$,
\begin{equation}
q(c'|c) = \prod_i q(Y_i|X_i).
\end{equation}
$I^*(X;Y)$ is an extension of the mutual information $I(X;Y)$ in the sense of the information transfer rate over a noisy channel $p(Y|X)$ when a mismatched distribution $q(Y|X)$ is used for decoding. 

The information transfer rate determines the amount of information that can be obtained from a message. The receiver obtains more information from a message when the information transfer rate increases. The mutual information $I$, which is equivalent to the maximal information transfer rate, determines the maximum amount of information that can be obtained by matched decoding. $I^*$, in contrast, determines the amount of information that can be obtained by a mismatched decoding.

\subsubsection*{Mathematical expression of $I^*$}
The amount of information for mismatched decoding can be evaluated using the following equation,
\begin{multline}
I^* (X^{t-\tau};X^t)= -\sum_{X^t}  p(X^t)  \log \sum_{X^{t-\tau}} p(X^{t-\tau}) q(X^t|X^{t-\tau})^{\beta} \\
+ \sum_{X^{t-\tau}, X^t} p(X^{t-\tau},X^t)  \log q(X^t|X^{t-\tau})^{\beta}, \label{eq:Istar}
\end{multline}
where $\beta$ is the value that maximizes $I^*$. The maximization of $I^*$ with respect to beta is performed by differentiating $I^*$ and solving the equation, $d I^*(\beta)/d \beta=0$. In general, the solution of the equation can be found using the standard gradient ascent method, because $I^*$ is a convex function with respect to $\beta$ \cite{Merhav1994, Latham2005}.

For comparison, the mutual information is given by
\begin{multline}
I(X^{t-\tau};X^t)= -\sum_{X^t}  p(X^t)  \log p(X^t) 
+   \sum_{X^{t-\tau}, X^t} p(X^{t-\tau},X^t)  \log p(X^t|X^{t-\tau}). \label{eq:MI}
\end{multline}
If a mismatched probability distribution $q(X^t|X^{t-\tau})$ is replaced by the actual distribution $p(X^t|X^{t-\tau})$ in Eq. \ref{eq:Istar}, the derivative of $I^*$ becomes 0 when $\beta=1$. By substituting $q=p$ and $\beta=1$ into Eq. \ref{eq:Istar}, one can check that $I^*$ is equal to $I$ in Eq. \ref{eq:MI}, as it should be. The amount of information for mismatched decoding, $I^*$, was first derived in the field of information theory as an extension of the mutual information in the case of mismatched decoding \cite{Merhav1994}. $I^*$ was first introduced into neuroscience in \cite{Latham2005} and was first applied to the analysis of neural data by \cite{Oizumi2010}. However, $I^*$ in the prior neuroscience application \cite{Latham2005,Oizumi2010} was quantified between stimuli and neural states, not between past and present states of a system, as described in the present study.

\subsubsection*{Analytical computation of $\Phi^*$ under the Gaussian assumption}
Assume that the probability distribution of neural states $\mathbf{x}$ is the Gaussian distribution,
\begin{equation}
p(\mathbf{x}) = \frac{1}{\left((2 \pi)^N |\Sigma(X)| \right)^{1/2}} \exp \left( -\frac{1}{2} (\mathbf{x}-\bar{\mathbf{x}})^T \Sigma(X)^{-1} (\mathbf{x}-\bar{\mathbf{x}}) \right).
\end{equation}
where $N$ is the number of variables in $\mathbf{x}$, $\bar{\mathbf{x}}$ is the mean value of $\mathbf{x}$, and $\Sigma(X)$ is the covariance matrix of $\mathbf{x}$. The Gaussian assumption allows us to analytically compute $\Phi^*$, which reduces substantially the costs for computing $\Phi^*$. When $X^{t-\tau}$ and $X^t$ are both multivariate Gaussian variables, the mutual information between $X^{t-\tau}$ and $X^t$, $I(X^{t-\tau};X^t)$, can be analytically computed as
\begin{equation}
I(X^{t-\tau};X^t)=\frac{1}{2}  \log \frac{|\Sigma(X^{t-\tau})|}{|\Sigma(X^{t-\tau}|X^t)|}, 
\end{equation}
where $\Sigma(X^{t-\tau}|X^t)$ is the covariance matrix of the conditional distribution, $p(X^{t-\tau}|X^t)$, which is expressed as
\begin{equation}
\Sigma(X^{t-\tau}|X^t)= \Sigma(X^{t-\tau})- \Sigma(X^{t-\tau},X^t) \Sigma(X^t)^{-1} \Sigma(X^{t-\tau},X^t)^T,
\end{equation}
where $\Sigma(X^{t-\tau},X^t)$ is the cross covariance matrix between $X^{t-\tau}$ and $X^t$, whose element $\Sigma(X^{t-\tau},X^t)_{ij}$ is given by ${\rm cov}(X^{t-\tau}_i, X^t_j)$.

Similarly, we can obtain the analytical expression of $I^*$ as follows:
\begin{equation}
I^*(\beta) = \frac{1}{2} {\rm Tr} \left(\Sigma(X^{t}) R \right) + \frac{1}{2} \log \left( |Q| |\Sigma(X^{t-\tau})| \right) - \frac{\beta N}{2}, \label{eq:Istargauss}
\end{equation}
where $\rm Tr$ stands for trace. $Q$ and $R$ are given by
\begin{equation}
Q = \Sigma(X^{t-\tau})^{-1} + \beta \Sigma_D(X^{t-\tau})^{-1} \Sigma_D(X^t,X^{t-\tau})^T \Sigma_D(X^t|X^{t-\tau})^{-1} \Sigma_D(X^t,X^{t-\tau}) \Sigma_D(X^{t-\tau})^{-1},
\end{equation}
\begin{multline}
R = \beta \Sigma_D(X^t|X^{t-\tau})^{-1} \\
- \beta^2 \Sigma_D(X^t|X^{t-\tau})^{-1 T} \Sigma_D(X^t,X^{t-\tau}) \Sigma_D(X^{t-\tau})^{-1} Q^{-1} \Sigma_D(X^{t-\tau})^{-1} \Sigma_D(X^t,X^{t-\tau})^T \Sigma_D(X^t|X^{t-\tau})^{-1},
\end{multline}
where $\Sigma_D(X^{t-\tau})$, $\Sigma_D(X^t,X^{t-\tau})$ and $\Sigma_D(X^t|X^{t-\tau})$ are diagonal block matrices. Each block matrix is a covariance matrix of each part, $\Sigma (M_i^{t-\tau})$, $\Sigma (M_i^{t},M_i^{t-\tau})$, and $\Sigma (M_i^t|M_i^{t-\tau})$ where $M_i$ is a subsystem. For example, $\Sigma_D(X^{t-\tau})$ is given by

\begin{equation}
\Sigma_D(X^{t-\tau}) = 
\left(
\begin{array}{cccc}
\Sigma (M_1^{t-\tau}) &  &  &  \\
 & \Sigma (M_2^{t-\tau}) &  & $\mbox{\huge 0}$ \\
$\mbox{\huge 0}$ & & \ddots & \\
&  & &  \Sigma (M_m^{t-\tau})
\end{array}
\right).
\end{equation}

The maximization of $I^*$ with respect to $\beta$ is performed by solving the equation $d I^*(\beta)/d \beta=0$. The derivative of $I^*(\beta)$ with respect to $\beta$ is given by
\begin{equation}
\frac{d I^*(\beta)}{d \beta} = \frac{1}{2} {\rm Tr} \left( \Sigma(X^{t}) \frac{d R}{d \beta} \right) + \frac{1}{2}  {\rm Tr} \left( Q^{-1} \frac{d Q}{d \beta} \right) - \frac{N}{2},
\end{equation}
where
\begin{multline}
\frac{d R}{d \beta} = \Sigma_D(X^t|X^{t-\tau})^{-1} \\
- 2 \beta  \Sigma_D(X^t|X^{t-\tau})^{-1 T} \Sigma_D(X^t,X^{t-\tau}) \Sigma_D(X^{t-\tau})^{-1} Q^{-1} \Sigma_D(X^{t-\tau})^{-1} \Sigma_D(X^t,X^{t-\tau})^T \Sigma_D(X^t|X^{t-\tau})^{-1} \\
-\beta^2 \Sigma_D(X^t|X^{t-\tau})^{-1 T} \Sigma_D(X^t,X^{t-\tau}) \Sigma_D(X^{t-\tau})^{-1} \frac{d Q^{-1}}{d \beta} \Sigma_D(X^{t-\tau})^{-1} \Sigma_D(X^t,X^{t-\tau})^T \Sigma_D(X^t|X^{t-\tau})^{-1},
\end{multline}
\begin{equation}
\frac{dQ}{d \beta} = \Sigma_D(X^{t-\tau})^{-1} \Sigma_D(X^t,X^{t-\tau})^T \Sigma_D(X^t|X^{t-\tau})^{-1} \Sigma_D(X^t,X^{t-\tau}) \Sigma_D(X^{t-\tau})^{-1},
\end{equation}
and
\begin{align}
\frac{d Q^{-1}}{d \beta} &= -Q^{-1} \frac{d Q}{d \beta} Q^{-1}, \\
&= - Q^{-1} \Sigma_D(X^{t-\tau})^{-1} \Sigma_D(X^t,X^{t-\tau})^T \Sigma_D(X^t|X^{t-\tau})^{-1} \Sigma_D(X^t,X^{t-\tau}) \Sigma_D(X^{t-\tau})^{-1} Q^{-1}.
\end{align}
Inspection of the above equations reveals that $d I^*(\beta)/d \beta=0$ is a quadratic equation with respect to $\beta$. Thus, $\beta$ can be analytically computed without resorting to numerical optimization such as gradient ascent.

% Do NOT remove this, even if you are not including acknowledgments.

\section*{Acknowledgments}
M.O. was supported by a Grant-in-Aid for Young Scientists (B) from the Ministry of Education, Culture, Sports, Science, and Technology of Japan (26870860). N.T. was supported by Precursory Research for Embryonic Science and Technology from Japan Science and Technology Agency (3630), Future Fellowship (FT120100619) and Discovery Project (DP130100194) from Australian Research Council.

% Either type in your references using
% \begin{thebibliography}{}
% \bibitem{}
% Text
% \end{thebibliography}
%
% OR
%
% Compile your BiBTeX database using our plos2009.bst
% style file and paste the contents of your .bbl file
% here.
% 

% \section*{Figure Legends}
% This section is for figure legends only, do not include
% graphics in your manuscript file.
%
%\begin{figure}
%\caption{
%{\bf Bold the first sentence.}  Rest of figure caption.  
%}
%\label{Figure_label}
%\end{figure}

% \section*{Tables}
% 
% See introductory notes if you wish to include sideways tables.
%
% NOTE: Please look over our table guidelines at http://www.plosone.org/static/figureGuidelines#tables to make sure that your tables meet our requirements. Certain types of spacing, cell merging, and other formatting tricks may have unintended results and will be returned for revision.
%
%\begin{table}[!ht]
%\caption{
%\bf{Table title}}
%\begin{tabular}{|c|c|c|}
%table information
%\end{tabular}
%\begin{flushleft}Table caption
%\end{flushleft}
%\label{tab:label}
% \end{table}

% \section*{Supporting Information Legends}
%
% Please enter your Supporting Information captions below in the following format:
%\item{\bf Figure SX. Enter mandatory title here.} Enter optional descriptive information here.
% 
%\begin{description}
%\item {\bf}
%\item {\bf}
%\end{description}

\end{document}

% Title must be 150 characters or less
\begin{flushleft}
{\Large
\textbf{Measuring integrated information from the decoding perspective}
}
% Insert Author names, affiliations and corresponding author email.
\\
Masafumi Oizumi$^{1, 2, \ast}$, 
Shun-ichi Amari$^{1}$,
Toru Yanagawa$^{1}$,
Naotaka Fujii$^{1}$,
Naotsugu Tsuchiya$^{2,3,\ast}$
\\
\bf{1} RIKEN Brain Science Institute, 2-1 Hirosawa, Wako, Saitama 351-0198, Japan
\\
\bf{2} Monash University, Clayton Campus, Victoria 3800, Australia
\\
\bf{3} Japan Science and Technology Agency, Japan
\\
$\ast$ E-mail: oizumi@brain.riken.jp, naotsu@gmail.com
\end{flushleft}

\section*{Supporting Information}

\subsection*{Equivalence of $\Phi^*$ with $\Phi$ under the assumption of maximum entropy distribution}
We show that $\Phi^*$ proposed in the present paper is equivalent to $\Phi$, proposed by \cite{Balduzzi2008} when the maximum entropy distribution is assumed for the past state as follows. 

First, we should note that when the maximum entropy distribution is assumed, the distribution of past states in the whole system can be decomposed into the product of the distribution of each part as
\begin{equation}
p(^{\max}X^{t-\tau}) = \prod_i p(^{\max}M^{t-\tau}_i).
\end{equation}
Bearing this in mind, we can compute $I^*$ as follows
\begin{align}
I^*(\beta) &= - \int dX^t p(X^t) \log \prod_i \int dM^{t-\tau}_i p(^{\max}M^{t-\tau}_i) p(M^t_i|M^{t-\tau}_i)^{\beta} \nonumber \\
&+ \beta \int dX^{t-\tau}  \int dX^t p(^{\max}X^{t-\tau})p(X^t|X^{t-\tau}) \log \prod_i p(M^t_i|M^{t-\tau}_i), \\
&= - \sum_i \int d M^t_i p(M^t_i) \log \int dM^{t-\tau}_i p(^{\max}M^{t-\tau}_i) p(M^t_i|M^{t-\tau}_i)^{\beta} - \beta \sum_i H(M^t_i|M^{t-\tau}_i).
\end{align}
To obtain $\beta$ that maximizes $I^*$, we differentiate $I^*(\beta)$ as
\begin{equation}
\frac{d I^*(\beta)}{d \beta} = - \sum_i \int dM^t_i \frac{p(M^t_i)}{r(M^t_i)} \frac{d r(M^t_i)}{d \beta} - \sum_i H(M^t_i|M^{t-\tau}_i). 
\end{equation}
where
\begin{equation}
r(M^t_i) = \int dM^{t-\tau}_i p(^{\max}M^{t-\tau}_i) p(M^t_i|M^{t-\tau}_i)^{\beta},
\end{equation}
\begin{equation}
\frac{d r(M^t_i)}{d \beta} = \int dM^{t-\tau}_i p(^{\max}M^{t-\tau}_i) p(M^t_i|M^{t-\tau}_i)^{\beta} \log p(M^t_i|M^{t-\tau}_i), 
\end{equation}
Substituting $\beta=1$ into $\frac{d I^*(\beta)}{d \beta}$, we obtain
\begin{align}
\frac{d I^*(\beta=1)}{d \beta} &= - \sum_i \int dM^t_i \int dM^{t-\tau}_i p^{\max} (M^{t-\tau}_i) p(M^t_i|M^{t-\tau}_i) \log p(M^t_i|M^{t-\tau}_i) - \sum_i H(M^t_i|^{\max} M^{t-\tau}_i), \\
&= 0,
\end{align}
We therefore find that $I^*(\beta)$ is maximized when $\beta=1$. Substituting $\beta=1$ into $I^*(\beta)$, we obtain the expression of $I^*$ as
\begin{align}
I^*(\beta=1) &= \sum_i H(M^t_i) - \sum_i H(M^t_i|^{\max} M^{t-\tau}_i), \\
&= \sum_i I(^{\max} M^{t-\tau}_i; M^t_i). \label{eq:equiv}
\end{align}
From Eq. \ref{eq:equiv}, we see that our measure is equivalent to the original measure when the maximum entropy distribution is used.

\subsection*{$\Phi^*$ is 0 when parts are perfectly correlated}
We show that $\Phi^*$ is 0 when parts are perfectly correlated. For simplicity, we consider a system cosisting of two units $M_1$ and $M_2$ and the mismatched decoder, $q(X^t|X^{t-\tau})=p(M^t_1|M^{t-\tau}_1) p(M^t_2|M^{t-\tau}_2)$. It is easy to generalize to the case of more than two units. When the two units are perfectly correlated, $M^{t-\tau}_1=M^{t-\tau}_2$ and $M^{t}_1=M^{t}_2$. In this case, the joint probability distribution can be written as $p(X^{t-\tau})=p(M^{t-\tau}_1)p(M^{t-\tau}_2|M^{t-\tau}_1)= p(M^{t-\tau}_1) \delta(M^{t-\tau}_2-M^{t-\tau}_1)$ where $\delta(x)$ is the Dirac delta function. The first term of $I^*$ can be calculated as follows.

\begin{align}
I^*(\beta) &= - \int dX^t p(M_1^t) \delta(M^{t}_2-M^{t}_1) \log \int dX^{t-\tau} p(M^{t-\tau}_1) \delta(M^{t-\tau}_2-M^{t-\tau}_1) \prod^2_{i=1} p(M^t_i|M^{t-\tau}_i)^{\beta} \nonumber \\
&+ \beta \int dX^{t-\tau}  \int dX^t p(X^{t-\tau},X^t) \log \prod^2_{i=1} p(M^t_i|M^{t-\tau}_i), \\
&= - \int d M^t_1 p(M^t_1) \log \int dM^{t-\tau}_1 p(M^{t-\tau}_1) p(M^t_1|M^{t-\tau}_1)^{2\beta} - 2 \beta H(M^t_1|M^{t-\tau}_1). \label{eq:perfectcor}
\end{align}

To obtain $\beta$ that maximizes $I^*$, we differentiate $I^*(\beta)$ as
\begin{equation}
\frac{d I^*(\beta)}{d \beta} = - \int dM^t_1 \frac{p(M^t_1)}{r(M^t_1)} \frac{d r(M^t_1)}{d \beta} - 2 H(M^t_1|M^{t-\tau}_1). 
\end{equation}
where
\begin{equation}
r(M^t_1) = \int dM^{t-\tau}_1 p(M^{t-\tau}_1) p(M^t_1|M^{t-\tau}_1)^{2\beta},
\end{equation}
\begin{equation}
\frac{d r(M^t_1)}{d \beta} = 2 \int dM^{t-\tau}_1 p(M^{t-\tau}_1) p(M^t_1|M^{t-\tau}_1)^{2\beta} \log p(M^t_1|M^{t-\tau}_1), 
\end{equation}

Substituting $\beta=1/2$ into $\frac{d I^*(\beta)}{d \beta}$, we obtain
\begin{align}
\frac{d I^*(\beta=1/2)}{d \beta} &= - 2 \int dM^t_1 \int dM^{t-\tau}_1 p(M^{t-\tau}_1) p(M^t_1|M^{t-\tau}_1) \log p(M^t_1|M^{t-\tau}_1) - 2 H(M^t_1|M^{t-\tau}_1), \\
&= 0,
\end{align}

We therefore find that $I^*(\beta)$ is maximized when $\beta=1/2$. Substituting $\beta=1/2$ into $I^*(\beta)$, we obtain the expression of $I^*$ as
\begin{align}
I^*(\beta=1/2) &= H(M^t_1) - H(M^t_1|M^{t-\tau}_1), \\
&= I(M^{t-\tau}_1; M^t_1). 
\end{align}

When $M_1$ and $M_2$ are perfectly correlated, mutual information in the whole system is just equal to the mutual information in each part, i.e., $I(X^{t-\tau}; X^t)=I(M^{t-\tau}_1; M^t_1)=I(M^{t-\tau}_2; M^t_2)$. Thus, $\Phi^*$ becomes 0. 
\begin{align}
\Phi^* &= I(X^{t-\tau}; X^t) - I^*(X^{t-\tau}; X^t), \\
&= I(M^{t-\tau}_1; M^t_1) - I(M^{t-\tau}_1; M^t_1), \\
&= 0.
\end{align}

\subsection*{Theoretical requirements for a measure of integrated information are not satisfied by previously proposed measures}

\begin{figure}[t!]
\renewcommand{\figurename}{Supplementary Figure}
\begin{center}
\centerline{\includegraphics[width=0.8\textwidth]{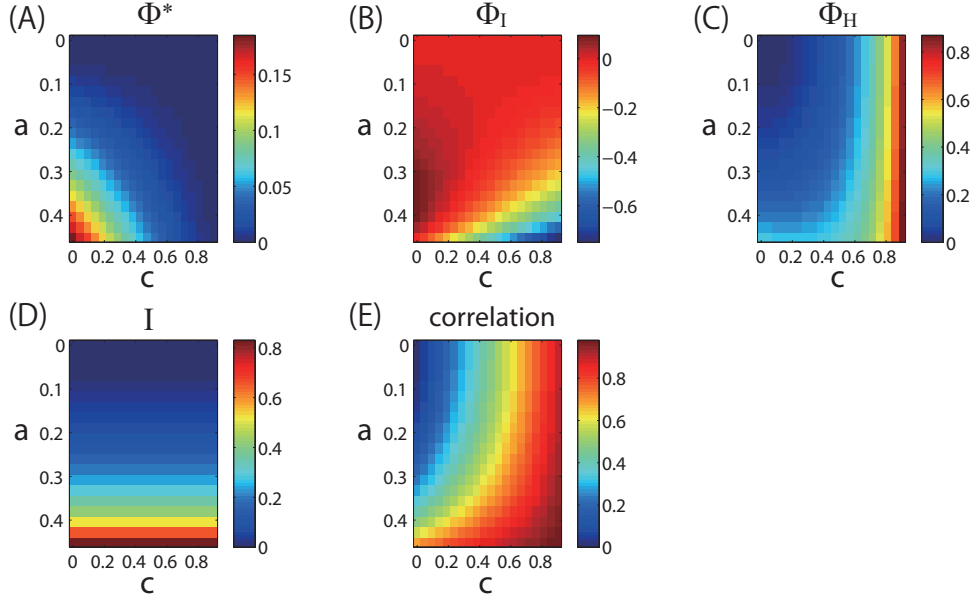}}
\caption{Behaviors of $\Phi^*$ (A), $\Phi_I$ (B), $\Phi_H$ (C), mutual information $I$ (D), and correlation (E) when the sterength of connections $a$ and the strength of noise correlation $c$ are both varied in a linear regression model (Eq. 12 in the main text). } 
\end{center}
\end{figure}

As we detailed in the main article, integrated information should be lower bounded by 0 and be upper bounded by the mutual information in the whole system $I$. In the main article, we used a simple linear regression model to demonstrate that $\Phi_I$ and $\Phi_H$ violate these bounds. Fig. S1 shows the behaviors of $\Phi^*$ (A), $\Phi_I$ (B), $\Phi_H$ (C), mutual information $I$ (D), and correlation coefficient between units (E) when the strength of connections $a$ and the strength of noise correlation $c$ are both varied in the same linear regression model as in the main article (Eq. 12). As we can see in Fig. S1, $\Phi_I$ goes negative when the degree of correlation is high and thus, it does not satisfy the lower bound. $\Phi_H$ is not 0 even when there is no information ($a=0$) and thus, it does not satisfy the upper bound. By comparing the panels (C) and (D), which are in the same color scale, we can see that $\Phi_H$ violates the upper bound when the noise correlation $c$ is high. $\Phi^*$ always satisfies both the lower bound and the upper bound and therefore, it can be considered as a proper measure of integrated information.